\documentclass[aps,prstab,twocolumn,groupedaddress,nofootinbib,floatfix]{revtex4}
\usepackage{epsfig, amsmath}
\newcommand{\eq}[1]{\mbox{Eq.~(\ref{#1})}}
\newcommand{\fig}[1]{\mbox{FIG.~\ref{#1}}}
\newcommand{\tab}[1]{\mbox{TABLE~\ref{#1}}}

\begin{document}

%\preprint{CERN/PS/OP 2002-???}
\preprint{physics/0202058}
%\preprint{to be submitted to Phys. Rev. ST-AB}

\title{Non-invasive single-bunch matching and emittance monitor}
\author{A. Jansson}
\email[]{Andreas.Jansson@cern.ch}
\affiliation{CERN, CH-1211 Geneva 23, Switzerland}

\date{\today}

\begin{abstract}
On-line monitoring of beam quality for high brightness beams is only 
possible using non-invasive instruments. For matching measurements,
very few such instruments are available. One candidate is a quadrupole pick-up.  
Therefore, a new type of quadrupole pick-up has been developed for the 
26 GeV Proton Synchrotron (PS) at CERN, and a measurement system consisting of 
two such pick-ups is now installed in this accelerator. 
Using the information from these pick-ups, it is possible
to determine both injection matching and emittance in the horizontal and vertical
planes, for each bunch separately.
This paper presents the measurement method and some of the results
from the first year of use, as well as comparisons with other 
measurement methods.
\end{abstract}



\maketitle

\section{Introduction and background}

A quadrupole pick-up is a non-invasive device that measures the quadrupole moment 
\begin{equation}
  \kappa=\sigma^2_x - \sigma^2_y + \bar{x}^2 - \bar{y}^2.
\label{e:quadmoment}
\end{equation}
of the transverse beam distribution.
Here, $\sigma_x$ and $\sigma_y$ are the r.m.s. 
beam dimensions in the horizontal and vertical directions, while 
$\bar{x}$ and $\bar{y}$ denote the beam position. 

The practical use of quadrupole pick-ups was pioneered at SLAC\cite{miller83}, where 
six such pick-ups, distributed along the linac, were used. The emittance and Twiss 
parameters of a passing bunch were obtained from the pick-up measurements 
by solving a matrix equation, derived from the known transfer matrices between pick-ups. 

In rings, the use of quadrupole pick-ups has largely focused on 
the frequency content of the raw signal. 
Beam width oscillations produce sidebands to the revolution frequency harmonics 
in the quadrupole signal, at a distance of 
twice the betatron frequency, %i.e. 
%$f_{\rm rev} (n\pm 2 q_{h,v})$, 
and
this can be used to detect injection mismatch. This was done at the
CERN Antiproton Accumulator, where the phase and amplitude of the detected sidebands 
were also used to find a proper correction, using an empirical response 
matrix\cite{chohan90}. However, this measurement was complicated by the fact 
that the same sidebands can be produced by position oscillations, which 
demanded that position oscillations were kept very small.

In this paper, the idea behind the SLAC method is applied and further developed 
for use in rings. 
The quadrupole pick-ups used for the measurements presented here were 
specially developed for the CERN PS and optimized to measure the quadrupole 
moment\cite{jansson00}. 

\begin{figure}[t]
   \includegraphics[width=\linewidth]{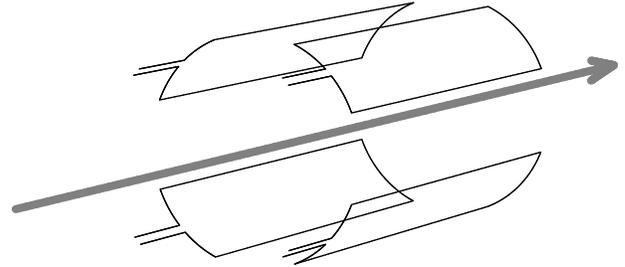}
    \caption{\label{f:0} Location of the induction loops. The arrow symbolises 
	the beam. For a centered, round beam no flux passes thru the loop
	and therefore no signal is induced. Therefore, the quadrupole signal 
	due to beam ellipticity is easy to detect.}
\end{figure}
They consist of four induction loops oriented to be sensitive to the
magnetic flux in the radial direction (see \fig{f:0}). 
Since the field from a centered round 
beam has a flux only in the azimuthal direction, only deviations from 
roundness or position induce a signal in the loops.
Therefore each loop is directly sensitive to the quadrupole moment, 
unlike previous instruments where the quadrupole moment was extracted
by detecting tiny differences between four large electrode signals.

Two pick-ups have been installed in consecutive straight sections of the 
machine\cite{jansson01}. The optical parameters at their locations
are given in \tab{t:1}. As shown later, it is crucial that the 
pick-ups be installed at locations with different ratios between horizontal 
and vertical beta value. 
The phase advance between pick-ups is also an important input parameter
in the data analysis. In order to minimize the dependence of this phase advance 
on the programmed machine tunes and the beam intensity (space charge detuning), 
the pick-ups were installed as close as possible to each other.

\begin{table}[tb]
\begin{center}
\caption{\label{t:1}Beta function values and horizontal dispersion 
at the pick-up locations. The horizontal and vertical phase advances 
between the two locations are also given. 
The pick-ups are installed in consecutive straight sections of the 
PS machine.}
\begin{tabular}{|c|c|c|c|c|c|}
\hline
Name & $\beta_x$ & $\beta_y$ & $D_x$ & $\Delta\mu_x$ & $\Delta\mu_y$\\ 
\hline
QPU~03 &  22.0 m & 12.5 m & 3.04 m & \raisebox{-\totalheight}{0.365} &
\raisebox{-\totalheight}{0.368} \\
QPU~04 & 12.6 m & 21.9 m & 2.30 m & &\\
\hline
\end{tabular}
\end{center}
\end{table}

The PS pick-ups provide both beam position and quadrupole moment information, 
with bunch-by-bunch resolution, over several hundred turns. 
Since the beam position is also measured, its contribution to the quadrupole 
moment can be subtracted, leaving only the beam-size related part, $\sigma_x^2-\sigma_y^2$.
Throughout the rest of this paper, when referring to $\kappa$, 
it will be assumed that this `artificial centering' has been performed, unless stated 
otherwise.

\begin{figure}[hbt]
   \includegraphics[width=1.05\linewidth]{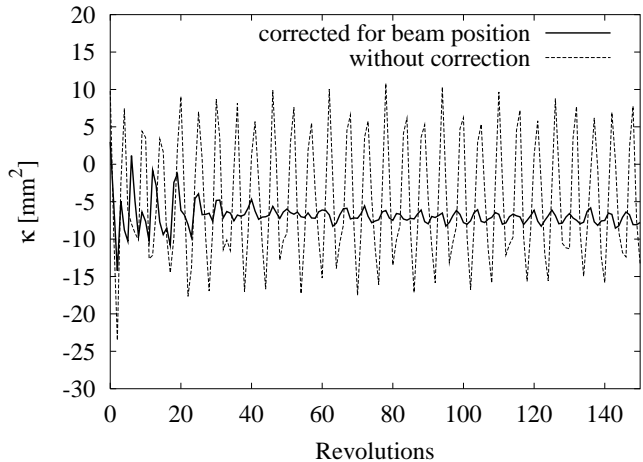}
    \caption{\label{f:1}Quadrupole moment $\kappa$ measured with a PS pick-up 
	immediately after injection, with and without correction for beam position.
	The initial beam-size oscillation is clearly visible in the corrected signal.  
	Note the fast decoherence of beam size oscillations, 
	due to direct space charge.}
\end{figure}

An example of a position-corrected measurement is shown in \fig{f:1}, where the 
usefulness of the correction is clear. 
The initial beam size oscillation due to injection mismatch is clearly 
visible in the corrected signal. Note that beam-size oscillations are sensitive 
to the direct space 
charge, which means that they have a larger tune-spread, and therefore decohere much 
faster than beam position oscillations. 
The difference in decoherence time between beam size and position oscillations is therefore 
a  rather  direct measure of the incoherent tune shift. 
The detuning of the quadrupole signal frequencies can also be used to measure the
incoherent tune shift, as has been done in the Low Energy Antiproton Ring (LEAR) at 
CERN\cite{chanel96}.

\section{Signal Acquisition and treatment}

At the input to the data acquisition system, located in a building 
next to the machine, the analogue signals from the pick-up have the form
\begin{eqnarray}
\Xi(t)      &=& Z_\Xi \; \kappa(t-t_\Xi) \; i(t-t_\Xi) \quad\;\; \text{quad. moment}\\
\Delta_x(t) &=& Z_{\Delta} \; \bar{x}(t-t_{\Delta_x}) \; i(t-t_{\Delta_x})
	\quad \text{hor. position}\\
\Delta_y(t) &=& Z_{\Delta} \; \bar{y}(t-t_{\Delta_y}) \; i(t-t_{\Delta_y})
 	\quad \text{ver. position}
\end{eqnarray}
where the $Z$s are the transfer impedances, $\kappa$, $\bar{x}$ and 
$\bar{y}$ are defined as before, and $i$ is the beam current.
In the PS, the beam current is not measured by the pick-up itself, 
so a separate beam current reference signal 
\begin{equation}
\Sigma(t)= Z_\Sigma \; i(t-t_\Sigma) \quad \text{sum signal}
\end{equation}
is taken from a nearby wall-current monitor. 
These analogue signals are sampled by digital oscilloscopes.
The digitized signals are then re-sampled\footnote{Eventually, 
the digital re-sampling will be replaced by analogue delay lines to 
improve the noise performance.}
to correct for the signal timing differences $t_\Xi$, $t_{\Delta_x}$, 
$t_{\Delta_x}$ and $t_\Sigma$. 
These are mainly due to cable length differences, and have been measured 
both with a synthetic signal and using the beam. 

The analysis of the data is made in a LabView program.
In order to resolve single bunches, the data is treated in the
time domain, considering each bunch passage separately. 
The first step in the analysis is to rid the signal of its intensity 
dependence, by normalizing to the measured beam current. The analysis is performed
in two different ways, depending on whether the position and quadrupole 
moment are expected to be constant or varying along the bunch.

\subsection{Position and size constant along bunch}

If there is no variation in position and size along the bunch, and one assumes 
that the quadrupole pick-up and the wall current 
monitor have the same frequency response, then the shape a given pulse  must be 
exactly the same in all signals (apart from a baseline offset and noise effects). 
The normalization problem then consists in determining 
the scaling factor between a pulse in the beam current signal and the 
corresponding pulse on the pick-up outputs.

To do this, time slices of about one RF period centered on the 
bunch are selected. Each selected slice is a vector of $N$ samples and, 
under the above assumption, corresponding slices are 
proportional to each other.
The quadrupole moment can therefore be found as the least
squares solution to an overdetermined matrix equation, 
which in the case of the quadrupole signal has the form
\begin{equation}
  \begin{pmatrix}    
    \Sigma_1 & 1\\   
    \Sigma_2 & 1\\
    \vdots & \vdots\\
    \Sigma_N & 1
  \end{pmatrix}\cdot
  \begin{pmatrix}    
    \kappa \\ 
    c
  \end{pmatrix}=\frac{Z_\Sigma}{Z_\kappa}
  \begin{pmatrix}    
    \Xi_1 \\      
    \Xi_2 \\
    \vdots \\
    \Xi_N 
  \end{pmatrix}.
\end{equation}
The constant $c$ depends on the base line difference and is not used.
The same calculation is performed for the position signals, and the 
position contribution to the quadrupole moment is then subtracted. 

An attractive feature of this method, apart from noise suppression,
is that the base line is automatically, and unambiguously, corrected for.
Differences in frequency response of the two instruments
could be corrected by filtering the signals, if these 
responses are known. However, such sophisticated corrections would 
enhance noise, and are not necessary in the PS.

\subsection{Position or size varying along bunch}

Sometimes, there can be a variation in oscillation amplitude and phase 
along the bunch.
At injection into the PS, there are two main causes for this
\begin{itemize}
\item
The injection kicker pulse is not perfectly flat, which causes a variation of initial
position along the bunch. The result is a fast position oscillation in those parts of the 
bunch that did not receive the correct kick.
\item
If the injected beam is longitudinally mismatched, the mismatched bunch
will rotate in the bucket with the synchrotron frequency, causing the 
bunch length to oscillate.
When the bunch is tilted in longitudinal phase space there is a 
correlation between energy and time, apparent as a variation of the
mean energy along the bunch. 
The degree of correlation varies as the bunch rotates, and at a 
position with non-zero dispersion this gives rise to a slow head-tail 
oscillation at twice the synchrotron frequency. Both PS pick-ups are 
installed in dispersive regions, and are therefore sensitive to this 
effect.
\item
There can also be a variation of the beam dimensions along the bunch, 
as discussed toward the end of this paper.
\end{itemize}
In these cases, the basic assumption behind the algorithm described in the
previous section is no longer valid.
In fact, if the position varies along the bunch, any algorithm that 
calculates the average position 
and quadrupole moment of the bunch will give an erroneous result. 
Since 
\begin{equation}
<x^2> \neq <x>^2
\end{equation}
one can not simply use the average bunch position in 
\eq{e:quadmoment} when correcting for the position.
The correction must be done point-by-point along the bunch. 
For this purpose, a second normalization algorithm is used, which first establishes and 
subtracts the base line, and then calculates the position
\begin{equation}
x(t)=\frac{Z_\Sigma\,}{Z_{\Delta_x}}\frac{\Delta_x(t)}{\Sigma(t)}
\end{equation}
as well as the quadrupole moment in each point. After this correction, 
an average beam quadrupole moment can be calculated, but it is
also possible to study variations of the beam size along the bunch.

\begin{figure}[hbt]
\includegraphics[width=1.05\linewidth]{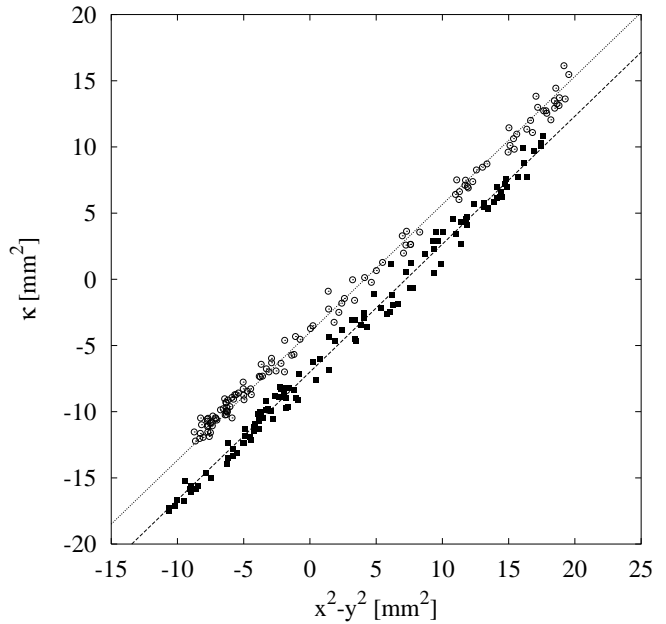}
    \caption{\label{f:2}Quadrupole moment (uncorrected) versus expected beam position contribution.
	The squares and circles represent measurements  made with the same pick-up on two 
	different beams.
	The slope of the line is the same in both cases, and  is very close to one (0.983).}
\end{figure}

\section{Beam-based calibration}
\subsection{Internal signal consistency}

One can take advantage of the position dependence of the quadrupole moment to make a consistency
check between the position and quadrupole moment measurement of the pick-up,
using data with large beam position oscillations but stable beam size. Since
the beam size oscillations damp away much faster than beam position oscillations, 
%due to the incoherent space charge tune shift, 
such data can easily be obtained at injection by an appropriate trigger delay. 
A plot of expected versus measured variation of the quadrupole moment
with beam position is shown in \fig{f:2}, showing a good agreement. 
This test can easily be automated, and is a good indicator of whether the beam 
position correction works well. 

\subsection{Comparison with wire-scanners}

\begin{figure*}[hbt]
 \includegraphics*[width=\linewidth]{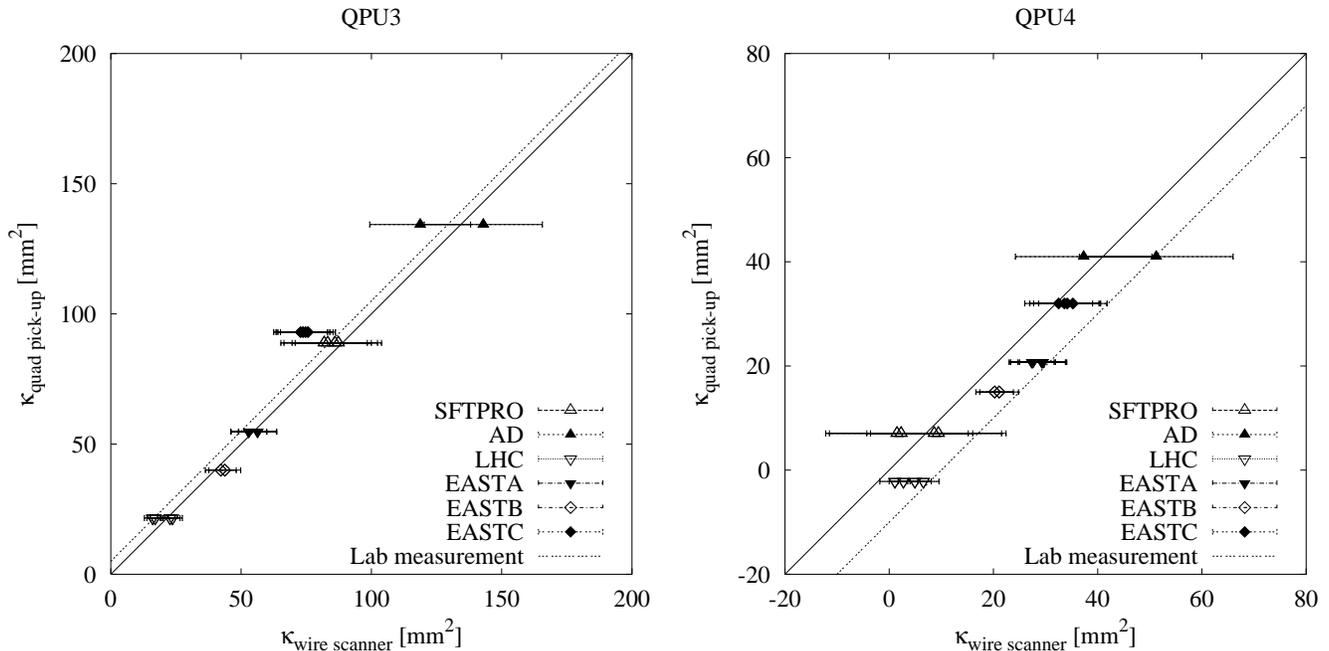}	 
 \caption{\label{f:32} Comparison between the measured value from the 
	two quadrupole pick-ups and the expected results calculated 
	from the emittances measured with the wire-scanners. The solid line
	is the ideal case, and the dotted line includes pick-up offsets measured in the 
	lab prior to installation. All possible ways of combining the wire-scanner measurements
	are displayed. Note that the cases where the two wire-scanner results are inconsistent 
	also are cases with large estimated systematic error.}
\end{figure*}

The standard method for emittance measurement on a circulating beam in the PS is 
the fast wire-scanner.
In order to test the calibration of the pick-ups, measurements were done
on several different stable beams, approximately 15 ms after injection. 
The quadrupole pick-up signal was acquired over 200 machine turns, at the same 
time as the wire traversed the beam. 
The comparative measurement was performed on all the operational beams available in 
the machine, with the exception of the very high intensity beams that saturate the 
pick-up amplifiers.
Thus there was a significant difference in both beam and machine parameters between
the different measurements. This was done in an attempt to randomize any
systematic errors. The beam parameters are given in \tab{t:2}, where the 
different beams have been tagged with their operational names.

\begin{table}[hbt]
\begin{center}
\caption{\label{t:2}
Parameters of beams used for comparative measurements. 
Emittances and momentum spread are 2$\sigma$ values.}
\begin{tabular}{|c|c|c|c|c|}
\hline
Name & $\epsilon_x$ & $\epsilon_y$ & $\sigma_{\rm p}$ & $I_{\rm bunch}$ \\
\hline
SFTPRO &  19 $\mu$m & 12 $\mu$m & 2.7 $\times 10^{-3}$ & $2.7\times 10^{12}$ \\
AD & 25 $\mu$m & 9 $\mu$m & 2.7 $\times 10^{-3}$ & $3.3\times 10^{12}$\\
LHC & 3 $\mu$m & 2.5 $\mu$m & 2.2 $\times 10^{-3}$ & $6.9\times 10^{11}$\\
EASTA & 8 $\mu$m & 1.4 $\mu$m & 2.5 $\times 10^{-3}$ & $1.4\times 10^{11}$ \\
EASTB & 7.5 $\mu$m & 1.4 $\mu$m & 1.6 $\times 10^{-3}$ & $8.6 \times 10^{10}$\\
EASTC & 12 $\mu$m & 3 $\mu$m & 2.4 $\times 10^{-3}$ & $4.2\times 10^{11}$\\
\hline
\end{tabular}
\end{center}
\end{table}

The r.m.s. variation in the measured quadrupole moments from turn to turn was of 
the order of  0.2-0.5~mm$^2$, depending on the beam 
intensity. Assuming that the beam size was perfectly stable, this gives an estimate 
of the single-turn resolution of the pick-up measurement. Also the wire-scanner 
measurements were stable, although for some beams there was a systematic disagreement 
between the two wire-scanners measuring in the same plane.

To compare the two instruments, the emittances measured with the wire-scanners were 
used to calculate the expected quadrupole moment at the locations of the pick-ups. 
The momentum spread required for both the wire scanner measurement and the subsequent 
calculation was obtained by a tomographic analysis of the bunch shape\cite{hancock99}.
The propagated systematic error was estimated on the assumption
that the wire-scanner accuracy is 5\% in emittance, the beta function at the 
pick-ups is known to 5\%, the dispersion to 10\% and the momentum spread to 3\%
accuracy. These estimates are rather optimistic, but give considerable propagated 
errors for certain measurement points. 
For simplicity, possible correlations between errors (e.g. beta function errors at 
different locations in the machine) were ignored, and all different error sources 
were added in quadrature.
To accentuate the cases with wire-scanner disagreement, each of the four different ways 
of combining the two horizontal and two vertical wire-scanners was calculated separately
and displayed as separate points. The result is shown in \fig{f:32}.

Overall, the measured data seem to indicate that the offsets are slightly smaller than 
measured in the lab, which could be explained by the fact that the pick-ups were 
dismantled in the lab to be moved to the machine. However, the effect is within the 
error-bar, and no strong conclusion can therefore be made. Moreover, the pick-ups have 
been dismantled and rebuilt in the lab, without effect on the measured offsets.

The point corresponding to the EASTC beam appears to disagree somewhat in both planes, 
although the effect is just about within the error-bar. There are a number of possible 
explanations for this:
\begin{itemize}
\item
The PS is operated in a time-sharing mode, where a so-called super-cycle containing
a certain number (usually 12) of beam cycles is repeated over and over again.
At the time of the measurement, the super-cycle contained several instances
of the EASTC beam, and it is known from experience that the position within
the super-cycle can affect the beam characteristics.
For this particular measurement, it is not guaranteed that the measurements 
with the two instruments were done on the same instance of the beam, whereas
for all other measurements there was either only one instance of the beam in the 
super-cycle, or the acquisition was locked to a certain instance. Some fluctuations
of the measured value were also observed.
\item
The EASTC beam has a large momentum spread and a horizontal tune close to an integer resonance. 
Theory indicates that the correction quadrupoles used to obtain this working point can perturb 
the dispersion function by more than 15\%\cite{carli01}, which would affect both the accuracy of 
the wire-scanner measurement, and the subsequent calculation of the expected quadrupole moment.
Studies of this effect are planned for the 2002 run.
\end{itemize}

The general conclusion from the measurement series is that the wire-scanner and quadrupole 
pick-up agree within the error bar. The systematic errors due to optics parameters make it impossible to 
detect with certainty any difference in pick-up behavior between the laboratory measurements
with a simulated beam, and the measurements on real beam in the machine. 
In order to calibrate the pick-ups accurately using the beam, the 
wire-scanners and the pick-up should be situated in the same straight section, 
which is excluded in the PS due to space limitations.

\subsection{Comparison with turn-by-turn profile measurement}

Comparative measurements of injection matching have been done using a 
SEM grid with a fast acquisition system\cite{raich01}, that can measure beam profiles 
turn-by-turn for a single bunch. This is a destructive device and can 
only be used in rare dedicated machine development sessions. 
It is also limited both in bandwidth and maximum beam intensity, 
and therefore it has not been possible to make a full systematic study on 
beams with different characteristics. Instead, a special beam 
was prepared, with low intensity to spare the grid, and long bunches
due to the bandwidth limitations.

\begin{figure}[hbt]
   \includegraphics*[width=\columnwidth]{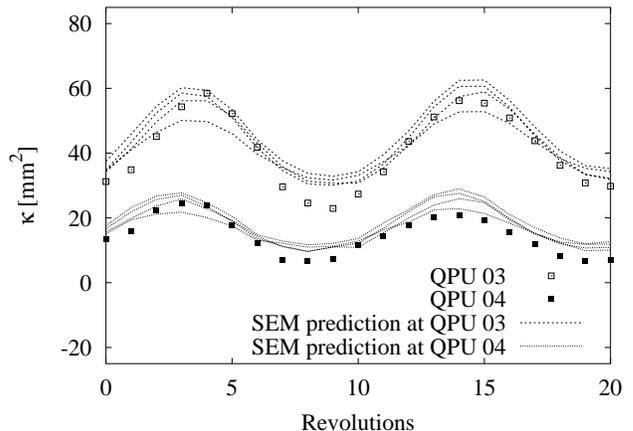}	 
 \caption{\label{f:22} Beam size oscillations at injection measured with the 
	quadrupole pick-ups and a turn-by-turn SEM grid. 
	The SEM-grid beam size data were used 
	to calculate the expected quadrupole
	moment at the pick-up locations. Beam position contributions and 
	known pick-up offsets have been subtracted from
	the quadrupole moments.}
\end{figure}

The SEM grid data was used to calculate the expected value of the quadrupole
moment at the pick-up locations, using the beta values, dispersion, and relative 
phase advance in Table~\ref{t:1}. 
The results are shown in \fig{f:22}, and show a rather good agreement with 
what was actually measured with the pick-ups. 
The small differences can be accounted for by systematic error sources,
i.e. the optical parameters used in the comparison.

\section{Emittance measurement}

When the circulating beam is stable, the quadrupole moments of a given bunch, 
as measured by the two pick-ups, are constant and given by
\begin{eqnarray}
\kappa_1&=& \epsilon_x \bar{\beta}_{x\,1}-\epsilon_y \bar{\beta}_{y\,1} + \bar{D}^2_{x\,1} \sigma_{\rm p}^2 \notag \\
\kappa_2&=& \epsilon_x \bar{\beta}_{x\,2}-\epsilon_y \bar{\beta}_{y\,2} + \bar{D}^2_{x\,2} \sigma_{\rm p}^2 
\label{e:eqnsyst}
\end{eqnarray}
where $\epsilon$ denotes the emittance, $\beta$ the beta value, 
$D$ the dispersion, and $\sigma_{\rm p}$ the relative momentum spread.
The bar over certain parameters indicate that these are properties of 
the lattice, to be distinguished from the corresponding beam properties
(typeset without the bar).

When the momentum spread is known, the system of equations can be solved for 
the emittances if 
\begin{equation}
\frac{\bar{\beta}_{x\,1}}{\bar{\beta}_{y\,1}} \neq \frac{\bar{\beta}_{x\,2}}{\bar{\beta}_{y\,2}}
\end{equation}
which explains the earlier statement about the requirement on the beta functions at the 
pick-up locations. If the ratio between horizontal and vertical beta function is significantly different
at the two locations, the equations are numerically stable. Thus measuring 
the emittance of a stable circulating beam with quadrupole pick-ups is in fact rather 
straightforward. 

Statistical errors due to random fluctuations in the measurement 
of $\kappa$ can, although they are usually small,  be reduced by
averaging over many consecutive beam passages.
The dominant errors are therefore systematic, coming from offsets in 
the pick-ups and errors in the beta functions, lattice dispersion and momentum spread. 
The pick-up offsets are, however, known from test bench measurements. 
Furthermore, by comparing the amplitude of position oscillations as measured by the two 
pick-ups, the ratios $\bar{\beta}_{x\,1}/\bar{\beta}_{x\,2}$ and 
$\bar{\beta}_{y\,1}/\bar{\beta}_{y\,2}$ can be 
determined.

\begin{figure}[hbt]
    \includegraphics[width=\linewidth]{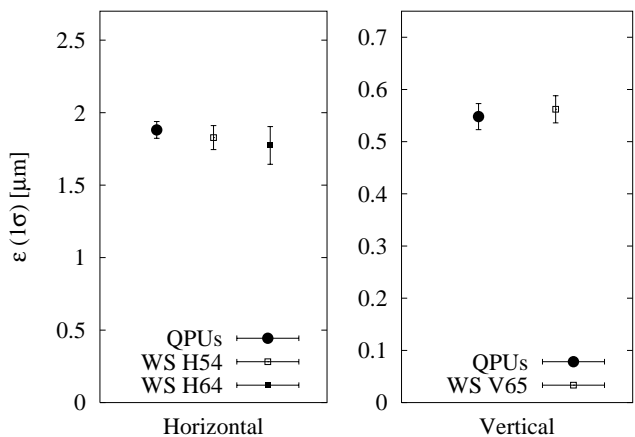}
    \caption{\label{f:23}Filamented emittance of a proton beam 
	measured with quadrupole pick-ups (QPU) and wire-scanner (WS). 
	There is a good agreement. The error bar is the standard deviation 
	for 10 measurements. Figure from \cite{jansson01pac}.}
\end{figure}

The main uncertainty is thus the absolute value of the beta function, as 
for almost any other emittance measurement (e.g. wire-scanner). 
The accuracy can therefore be expected to be comparable to that of a 
wire-scanner. An emittance measurement using the pick-up system 
is shown in \fig{f:23}, and compares well with wire-scanner 
results.

Note that with three pick-ups, suitably located, the momentum spread could also be measured.

\section{Matching measurement}

Even though quadrupole pick-ups can be used to measure filamented emittance, the main 
reason for installing such instruments in the machine is to be able to
measure betatron and dispersion matching at injection, as no other instrument (apart
from the destructive SEM-grid) is able to do this. One would like not only to 
detect mismatch, but also to quantify the injection error in order to be able to correct it.

\subsection{Matrix inversion method}

To determine the parameters of the injected beam, the SLAC method\cite{miller83}
based on matrix inversion could be directly applied, since the quadrupole moment is 
measured on a single-pass basis. 
An advantage when performing this measurement in a ring, as compared to measuring in 
a linac, is that each pick-up can be used several times on the same bunch. Therefore 
it is enough to use two pick-ups instead of six, which reduces both the hardware cost and 
the systematic error sources. It is also straightforward to improve on 
statistics by increasing the number of measured turns, thereby reducing noise. 
Another advantage in a ring is that the periodic boundary conditions reduce the 
number of parameters needed to calculate the matrix. Many of these parameters 
(tunes, phase advance between pick-ups, ratios between beta function values) 
can also be easily measured, which means that the matrix can be experimentally 
verified.
%
%The tunes and the pick-up phase separations can be obtained from the frequency 
%and the phase difference of beam position oscillations in the two planes. 
%Similarly, as pointed out previously, in each plane the ratio 
%of beta functions at the two pick-ups can be determined. 

However, the matrix method was developed for a linac, and does
not take full advantage of the properties of a ring. Also, it does not
include dispersion effects, and it is necessary to make assumptions 
on the space charge detuning when calculating the matrix.

\subsection{Parametric fit method}

In a ring, the turn by turn evolution of the beam envelope, and therefore the quadrupole 
moment, can be expressed in a rather simple analytical formula. Expanded in terms of the
optical parameters, the quadrupole moment of a beam is given by
\begin{equation}
\kappa= \sigma_x^2-\sigma_y^2 %+x^2 -y^2 
=\varepsilon_x \beta_{x} - \varepsilon_y \beta_{y} +
\sigma_p^2 D_x^2 - \sigma_p^2 D_y^2
\end{equation}
assuming linear optics with no coupling between planes (note that there are no 
bars, i.e. the optical parameters here refer to the beam).
If the beam is initially  mismatched in terms of Twiss functions or dispersion,
the value of $\kappa$ will vary with the number of revolutions $n$
performed  as\cite{janssondiss}
\begin{multline}
\kappa_n = \bar{\beta}_x (\varepsilon_x+ \Delta \varepsilon_x) - 
\bar{\beta}_y (\varepsilon_y+\Delta \varepsilon_y) 
+ \bar{D}^2_x\,\sigma_{\rm p}^2 % +\bar{x}^2-\bar{y}^2 
\\
+\bar{\beta}_x \varepsilon_x \delta_{\beta_x} \cos(2\nu_x n- \phi_{\beta_x})
+\bar{\beta}_x \sigma_{\rm p}^2 \delta_{D_x}^2 \cos(2\nu_x n- 2\phi_{D_x})\\
-\bar{\beta}_y \varepsilon_y \delta_{\beta_y} \cos(2\nu_y n- \phi_{\beta_y})
-\bar{\beta}_y \sigma_{\rm p}^2 \delta_{D_y}^2 \cos(2\nu_y n- 2\phi_{D_y})\\
+\sqrt{\bar{\beta}_x}\sigma_{\rm p}^2 \bar{D}_x\,\delta_{D_x} \cos(\nu_x n-\phi_{D_x})
\label{e:quadsig}
\end{multline}
%Here, barred parameters refer to properties of the lattice, 
Here, $\nu_x$ and $\nu_y$ are the fractional tunes expressed in radians, and
$\Delta\epsilon$ denotes the emittance increase caused by the mismatch. 
%$2 \pi q_{x,y}$.
%
%The terms have been ordered so that each line corresponds to one oscillation frequency. 
%
The first line contains constant terms, 
and also gives the steady state value that will be reached when the oscillating components 
have damped away. 

The two middle lines of \eq{e:quadsig} are signal components at twice the horizontal 
and vertical betatron frequencies. They arise from both dispersion and betatron mismatch. 
The betatron mismatch is parametrized by
\begin{equation}
\vec{\delta}_{\beta_x}=
\begin{pmatrix}
\frac{\beta_x}{\bar{\beta}_x} - 
\frac{\bar{\beta}_x \gamma_x + \bar{\gamma}_x \beta_x - 2 \bar{\alpha}_x\alpha_x}{2}\\
\frac{\bar{\alpha}_x \beta_x- \alpha_x\bar{\beta}_x}{\bar{\beta}_x}
\end{pmatrix}
\approx
\begin{pmatrix}
\frac{\Delta\beta_x}{\bar{\beta}_x} \\
\bar{\alpha}_x \frac{\Delta \beta_x}{\bar{\beta}_x} -\Delta\alpha_x 
\end{pmatrix}
\end{equation}
where the last approximation is valid for small mismatch. Here, the shorthand
notation $\Delta\beta = \beta-\bar{\beta}$ and  $\Delta\alpha = \alpha - \bar{\alpha}$ is 
used for the difference between lattice and beam value.

The fourth line of 
\eq{e:quadsig} is a signal at the 
horizontal betatron frequency, which is due to dispersion matching. This mismatch is 
parametrized by the vector
\begin{equation}
\vec{\delta}_{D_x}=
\begin{pmatrix}\frac{\Delta D_x}{\sqrt{\bar{\beta}_x}}\\
\frac{\bar{\beta}_x \Delta D'_x + \bar{\alpha}_x \Delta D_x}{\sqrt{\bar{\beta}_x}} 
\end{pmatrix}
\end{equation}
where, again, shorthand notation ($\Delta D = D -\bar{D}$ and $\Delta D' = D' -\bar{D}'$)
is used.
There is no corresponding signal at the vertical betatron frequency due to the 
absence of vertical lattice dispersion. Therefore, it is not possible to distinguish 
vertical dispersion mismatch from vertical betatron mismatch by studying the quadrupole 
signal. However, one does not usually expect a large vertical dispersion mismatch.

The steady state (filamented) emittance is given by
\begin{multline}
  \varepsilon_x + \Delta \varepsilon_x = 
  \varepsilon_x \frac{1}{2}\left(\bar{\beta}_x \gamma_x + \bar{\gamma}_x \beta_x - 
    2 \bar{\alpha}_x \alpha_x\right) +\\
  +\sigma_{\rm p}^2 \frac{(\Delta D_x)^2 + (\bar{\beta}_x \Delta D'_x +\bar{\alpha}_x \Delta D_x)^2}{\bar{\beta}_x} \approx\\
\approx  \varepsilon_x +
  \varepsilon_x \frac{|\vec{\delta}_{\beta_x}|^2}{2} 
  +\sigma^2_{\rm p} \frac{|\vec{\delta}_{D_x}|^2}{2} 
\end{multline}
where, again, the last approximation is valid for small betatron 
mismatch\footnote{There is also a contribution to the emittance increase 
due to injection miss-steering that is not included here, since normally 
coherent dipole oscillations filament much slower than quadrupole oscillations, 
and the beam position contribution is subtracted from the signal.}.

\begin{figure}[t]
   \includegraphics[width=1.05\linewidth]{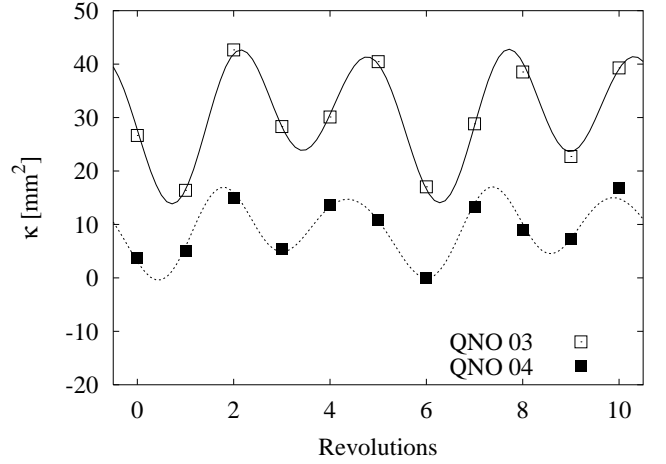}	 
    \caption{\label{f:3}Theoretical expression for the quadrupole moment fitted to 
	measured data. Here, seven turns (14 data points) were used to 
	 determine 10 free parameters (emittances, betatron and dispersion 
	mismatches, and the tunes), but there is a relatively good match also for the 
	subsequent turns. The measured detuning of the beam width oscillation frequencies 
	were quite significant,	$\Delta Q_h =0.01$ and $\Delta Q_v=0.05$ (as compared to 
the tunes measured from position oscillations). }
\end{figure}

By fitting the above function to the data, the injected emittances, the betatron
mismatches in both planes, and the horizontal dispersion mismatch are directly obtained.
The tunes can also be free parameters in the fit, which automatically estimates and corrects
for space charge detuning. An example of a fit to measured data is shown in \fig{f:3}.
A requirement for a good fit convergence is, as when measuring filamented emittance, that
the ratio between beta functions should be different at the pick-up locations. Also, the
tunes must be such that enough independent data points are obtained. In other words,
if the quadrupole signal is repetitive, it must have a period larger than the minimum 
number of turns required for the fit.
In the PS, this means that the working point $Q_h=Q_v=6.25$, which is close to the bare
tune, should be avoided. The fit result is also less stable in the vicinity of this 
working point, and when the tune in only one of the planes is close to 6.25.
With two pick-ups, at least five machine turns (10 data points) are required for the fit, 
if the tunes are also free parameters. 
Some more turns can be used to check the error, but the maximum number of turns is limited 
by decoherence, as discussed below.

Note that since the beam size oscillations due to dispersion mismatch are also 
detuned by space charge, measuring the dispersion component separately (by changing the energy 
of the beam and measuring the coherent response) would result in an accumulated phase 
error in the dispersion term.

\subsection{The effect of decoherence}

\begin{figure}[htb]
   \includegraphics[width=1.05\linewidth]{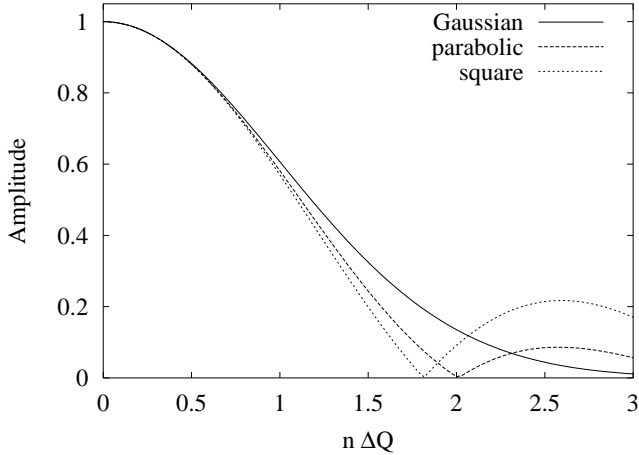}	 
    \caption{\label{f:35} Evolution of coherent amplitude during decoherence, for three
	different momentum (tune) distributions. Here, $\Delta Q$ is the r.m.s. tune spread and $n$ is the number of revolutions.}
\end{figure}

The fit function above does not include the effect of decoherence (damping) of the 
beam width oscillations.
Fortunately, due to the physics of the decoherence process, the decay of the oscillation 
amplitudes is not exponential as for many other damping phenomena.
If the beam is approximated by an ensemble of harmonic oscillators with a tune
distribution $\rho(\Delta Q)$ and an average tune Q, its coherent 
response to an initial displacement is
\begin{equation}
x(s)
=
e^{i 2\pi Q s}
\underset{A(s)}{
\underbrace{A_0 \int_{-\infty}^{\infty} e^{i 2\pi \Delta Q s} \rho(\Delta Q) \;d(\Delta Q)}}
\end{equation}
and the derivative of the amplitude function
\begin{equation}
 \frac{\partial A}{\partial s} 
\propto
\int_{-\infty}^{\infty} e^{i 2\pi \Delta Q s}\; \Delta Q\;\rho(\Delta Q) \;d(\Delta Q)
\end{equation}
is zero at $s=0$, i.e. initially the amplitude is unchanged by the 
decoherence process. 
A plot of the amplitude versus time for some tune distributions is
given in \fig{f:35}, showing that the initial behavior is also largely 
independent of the distribution.

In reality, the tune of each individual particle is changing with time (e.g. due to synchrotron 
motion), and therefore the decoherence pattern is more complicated. However, synchrotron motion
is negligible for the first few turns. Therefore, data analysis is greatly simplified and accuracy 
is improved, if one limits the number of turns to a rather small value. This also demonstrates an 
advantage of the fit method over an FFT analysis of the signals, since an FFT needs many points 
to achieve good frequency resolution.

\subsection{Measurement results}

\begin{figure*}[htb]
   \includegraphics[width=\linewidth]{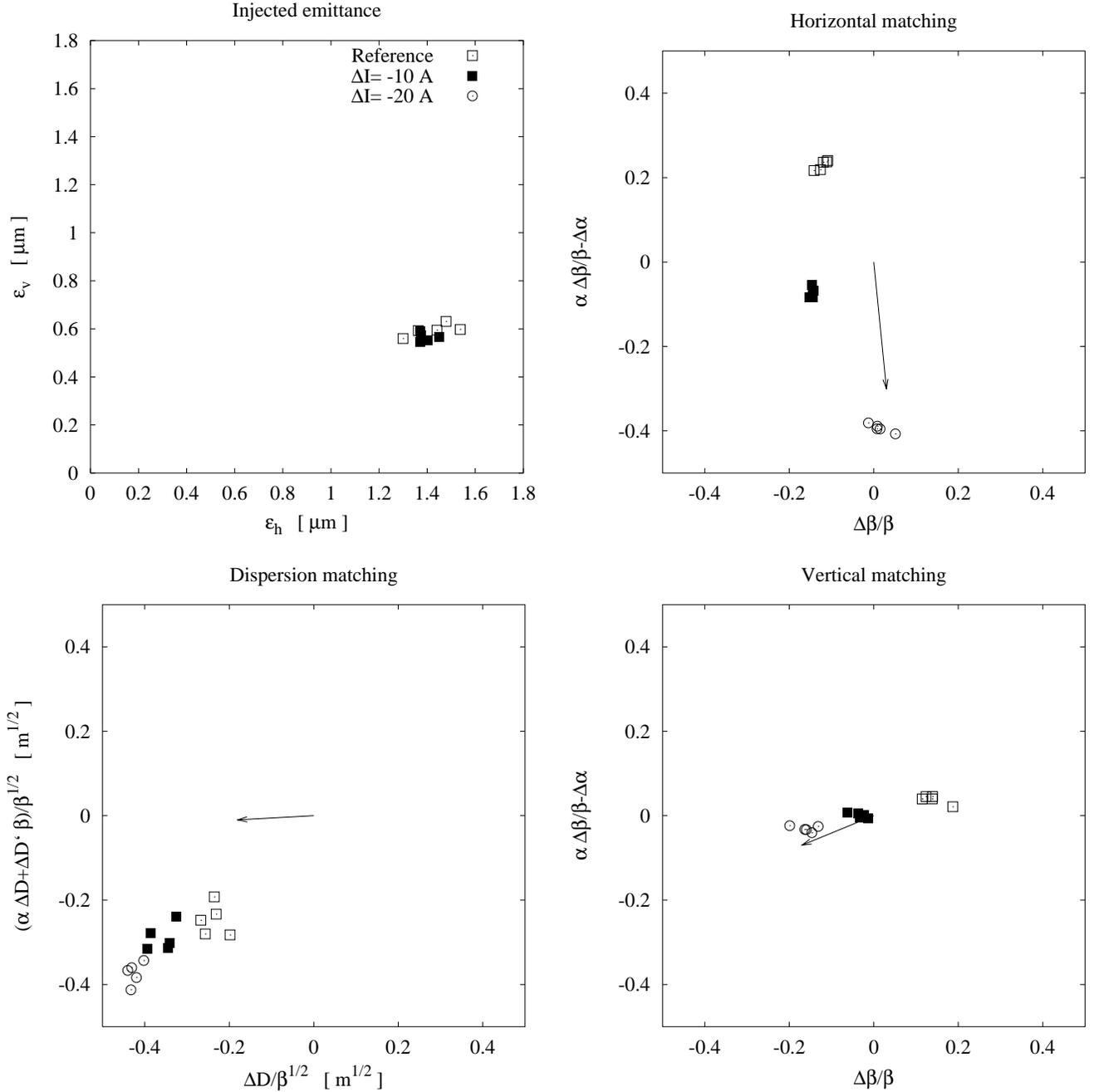}	 
 \caption{\label{f:33} Injected emittance, betatron and dispersion mismatch vectors for three 
	different settings of a transfer line quadrupole. Note the large dispersion mismatch. 
	The vectors illustrate the variation in mismatch that is expected 
	for a correction of -10A (calculated from beam optics theory). There is a good agreement 
	between expected and measured behavior, indicating that the measurement works well.}
\end{figure*}

To test the injection matching measurement, a series of measurements was done with different
settings of some focusing elements of the PS injection line. An example of such a measurement 
is shown in \fig{f:33}, where a quadrupole was changed in steps of $10\;$A, and the resulting 
variation of the fit parameters recorded. 
The variation of the different error vectors expected from beam optics theory is also shown, 
and there is a rather good agreement, both in direction and magnitude of the changes.
The injected emittances are unchanged, as expected.

By using the theoretical response matrix for dispersion
and betatron matching, a proper correction to the measured error can be 
calculated\cite{giovannozzi98}.
So far, actual corrections of the measured mismatches have not been made, since the dominant 
error (the dispersion mismatch) can not be corrected without a complete change of optics of 
the entire line. Studies for a new dispersion matched optics are underway.

While the dispersion mismatch is large for all beams, due to the transfer line design,
the level of betatron mismatch varies between different operational beams. 
Most high 
intensity beams measured were observed to be fairly well matched, whereas some lower 
intensity beams had a significant mismatch. This might be explained by the fact that 
mismatch is likely to cause losses for aperture limited beams, and therefore the process of 
intensity optimization leads to well-matched beams, although the mismatch is never 
directly measured. This indirect matching mechanism is absent for the future 
bright LHC beam, and it can therefore be expected to develop a relatively large mismatch 
if not continuously monitored and corrected. 

\section{Measurement within the bunch}

\begin{figure}
   \includegraphics*[width=\columnwidth]{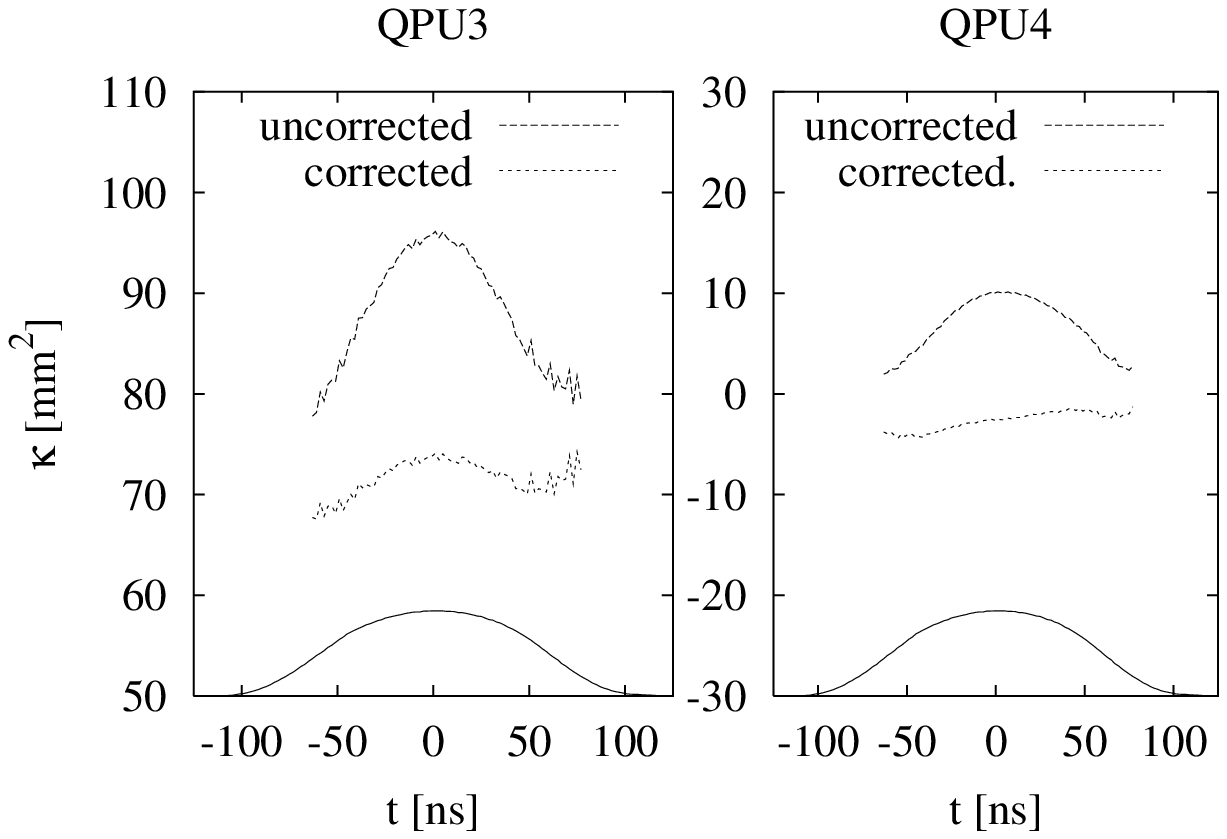}	 
 \caption{\label{f:4} Quadrupole moment as a function of position within the bunch, with and
	without correction for dispersive contribution. The bunch shape is also indicated 
	(solid line). }
\end{figure}

As mentioned earlier, the transverse mean position can sometimes
vary along the bunch. However, in some cases, also the beam size itself
varies along the bunch. This is notably the case for high 
intensity beams that are highly non-Gaussian.
For a Gaussian beam distribution, the transverse
bunch width is constant along the bunch. This is because the 
multi-dimensional Gaussian is just a product of one-dimensional 
Gaussians.
However, for a parabolic beam this is no longer true, as may be easily 
verified analytically.
With the pick-ups, it is possible to measure the quadrupole moment 
as a function of position within the bunch.  The measurement is good 
over most of the bunch, but naturally gets very noisy and prone to 
systematic errors in the head and tail, since these regions are sparsely populated.
A measurement made on a stable beam is shown in \fig{f:4}. 
The plot also shows the same measurement with the dispersion 
contribution subtracted\footnote{The momentum spread as a function of position 
within the bunch was 
obtained from a tomographic analysis\cite{hancock99} of the bunch 
shape data.}, indicating that the variation of beam size
along the bunch is mainly due to variations in momentum spread.
This fits with the fact that the longitudinal bunch distribution is usually 
non-Gaussian.
Applying the methods discussed earlier on the dispersion corrected data, it 
is also possible to calculate the emittance variations along the bunch.

\section{Conclusions}

The quadrupole pick-ups recently built and installed in the PS machine
have been evaluated in a series of measurements. 
These pick-ups measure both injection matching and emittance for a single, selected 
bunch in the machine. The measurement can be made parasitically, without perturbing the beam,
because the devices are non-intercepting.

Comparison with other instruments in the machine show good agreement. 
All observed deviations are within the estimated systematic error bars. 
The systematic errors come mainly from 
the imperfect knowledge of beta value and dispersion needed to evaluate 
the data. Systematic errors are indeed expected to dominate the 
total error in the quadrupole pick-up measurement, as 
is the case for most emittance measurement devices.

For matching applications, the pick-ups can be used to determine phase 
and amplitude of horizontal and vertical betatron mismatch, as well as
horizontal dispersion mismatch. This analysis can be done individually
on each injected bunch. Since the mismatch is detected as an
oscillation, the effect of systematic errors (e.g. pick-up offsets) 
is not very important.
 
As emittance measurement devices, the pick-ups have some interesting 
properties. The single turn resolution makes it possible to measure and 
follow the evolution of the emittance over many turns (limited only by 
acquisition memory). When measuring filamented emittance, it possible
to reduce the effect of noise by averaging over many turns, and also 
to check that the beam is stable during the measurement, something 
that is assumed but not actually verified during a wire scanner 
measurement.
More important, the pick-ups have no moving parts that wear out, as is 
the case for a wire-scanner. This makes it possible to create a watchdog 
application to monitor the evolution of the emittances pulse by pulse over 
a long period. In such an application, systematic errors are again of lesser 
importance, since variations rather than absolute values are sought.

The pick-ups can also be used
to study variations of the emittance along the bunch, 
although this may be mainly of academic interest.

\begin{acknowledgments}
The author would like to thank
D.J.~Williams and L.~S{\o}by for their support and important contributions 
to the pick-up hardware; H.~Koziol for reading and commenting on an early draft 
to this paper; U.~Raich and C.~Carli for contributing to the  
turn-by-turn SEM-grid data acquisition and analysis.
\end{acknowledgments}

\bibliography{paper}
\end{document}